\documentclass[10pt,journal,compsoc]{IEEEtran}

\usepackage{enumitem}

\usepackage{graphicx}
\usepackage{float}
\usepackage{array}
\usepackage{tabularx}
\usepackage{multirow}
\usepackage{amsmath}
\usepackage{amssymb}
\usepackage{threeparttable}
\usepackage{setspace}
\usepackage{ulem}
\usepackage{color}
\usepackage{colortbl}
\usepackage{xcolor}
\usepackage{soul}
\usepackage{booktabs}
\usepackage{picins}

\usepackage{algorithm} 
\usepackage{algpseudocode}
\usepackage{hyperref}
\usepackage{caption}

\definecolor{mygray}{gray}{.9}
\definecolor{mypink}{rgb}{.99,.91,.95}
\definecolor{mycyan}{cmyk}{.3,0,0,0}
\definecolor{myyellow}{RGB}{255,230,204}
\definecolor{mybule}{RGB}{218,232,252}
\definecolor{mygreen}{RGB}{213,232,212}
\definecolor{titleColor}{RGB}{102,102,102}

\definecolor{5007}{RGB}{127,127,127}

\title{AI Chain on Large Language Model for Unsupervised Control Flow Graph Generation for Statically-Typed Partial Code}
\author{Qing~Huang,
        Zhou~Zou,
        Zhenchang~Xing,
        Zhengkang~Zuo,
        Xiwei~Xu,
        Qinghua~Lu
\IEEEcompsocitemizethanks{\IEEEcompsocthanksitem Q. Huang, Z. Zou, Z. Zuo are with School of Computer Information Engineering, Jiangxi Normal University, China.\protect
\IEEEcompsocthanksitem Q. Huang and Z. Zou are co-first authors, Z. Zuo is the corresponding author(zuo803@jxnu.edu.cn).
\IEEEcompsocthanksitem Z. Xing, X. Xu and Q. Lu are with the CSIRO's Data61, Australia. 
}% <-this % stops an unwanted space
}
\begin{document}
\IEEEtitleabstractindextext{%
\begin{abstract}
Control Flow Graphs (CFGs) are essential for visualizing, understanding and analyzing program behavior. For statically-typed programming language like Java, developers obtain CFGs by using bytecode-based methods for compilable code and  Abstract Syntax Tree (AST)-based methods for partially uncompilable code. However, explicit syntax errors during AST construction and implicit semantic errors caused by bad coding practices can lead to behavioral loss and deviation of CFGs.
To address the issue, we propose a novel approach that leverages the error-tolerant and understanding ability of pre-trained Large Language Models (LLMs) to generate CFGs. Our approach involves a Chain of Thought (CoT) with four steps: structure hierarchy extraction, nested code block extraction, CFG generation of nested code blocks, and fusion of all nested code blocks' CFGs. To address the limitations of the original CoT's single-prompt approach (i.e., completing all steps in a single generative pass), which can result in an ``epic'' prompt with hard-to-control behavior and error accumulation, we break down the CoT into an AI chain with explicit sub-steps. Each sub-step corresponds to a separate AI-unit, with an effective prompt assigned to each unit for interacting with LLMs to accomplish a specific purpose.
Our experiments confirmed that our method outperforms existing CFG tools in terms of node and edge coverage, especially for incomplete or erroneous code. 
We also conducted an ablation experiment and confirmed the effectiveness of AI chain design principles: Hierarchical Task Breakdown, Unit Composition, and Mix of AI Units and Non-AI Units.
Our work opens up new possibilities for building foundational software engineering tools based on LLMs, as opposed to traditional program analysis methods.
\end{abstract}

}

\maketitle

% \vspace{-2mm}
% \section{INTRODUCTION}
\IEEEraisesectionheading{\section{Introduction}}
\label{introduction}
% Diverse, Guided, Extensible and explainable
%variety,interpretability,guidance,extensibility
%straighten out the relationships between these APIs in an understandable manner

\IEEEPARstart{T}{he} Control Flow Graph (CFG) serves as a cornerstone in software engineering, illustrating program behavior by showcasing statement sequences and the conditions governing their execution order~\cite{allen1970control}. 
As a graphical representation of program behavior, CFG plays a crucial role in numerous software engineering tasks, including code search~\cite{guo2020graphcodebert, chen2019capturing}, code clone detection~\cite{Wang2020DetectingCC, hu2018deep, wei2017supervised}, and code classification~\cite{wang2020modular, zhang2019novel}. 
These applications contribute to enhanced code quality and software performance, emphasizing the essential role of CFG within the software engineering realm.

When programming in statically-typed programming language such as Java, developers usually use the bytecode-based method~\cite{WALA, Soot} to generate CFGs from the compiled bytecode as it provides an optimized and simplified representation of the program behavior. 
However, if the source code is incomplete or uncompilable, developers may use the Abstract Syntax Tree (AST)-based approach~\cite{pawlak2016spoon} to generate CFGs directly from the source code. 
While the AST-based approach reveals the code's structure and control flow, explicit syntax errors during AST construction can result in behavioral loss, where some nodes/edges are missing from the CFG. 
Even after correcting syntax errors and compiling the code, implicit semantic errors caused by bad coding practices~\cite{gong2015dlint} can still lead to behaviorally-deviating CFGs from both bytecode- and AST-based approaches, referred to as behavioral deviation.

\begin{figure*}[t]
    \centering
    \includegraphics[width=0.95\textwidth]{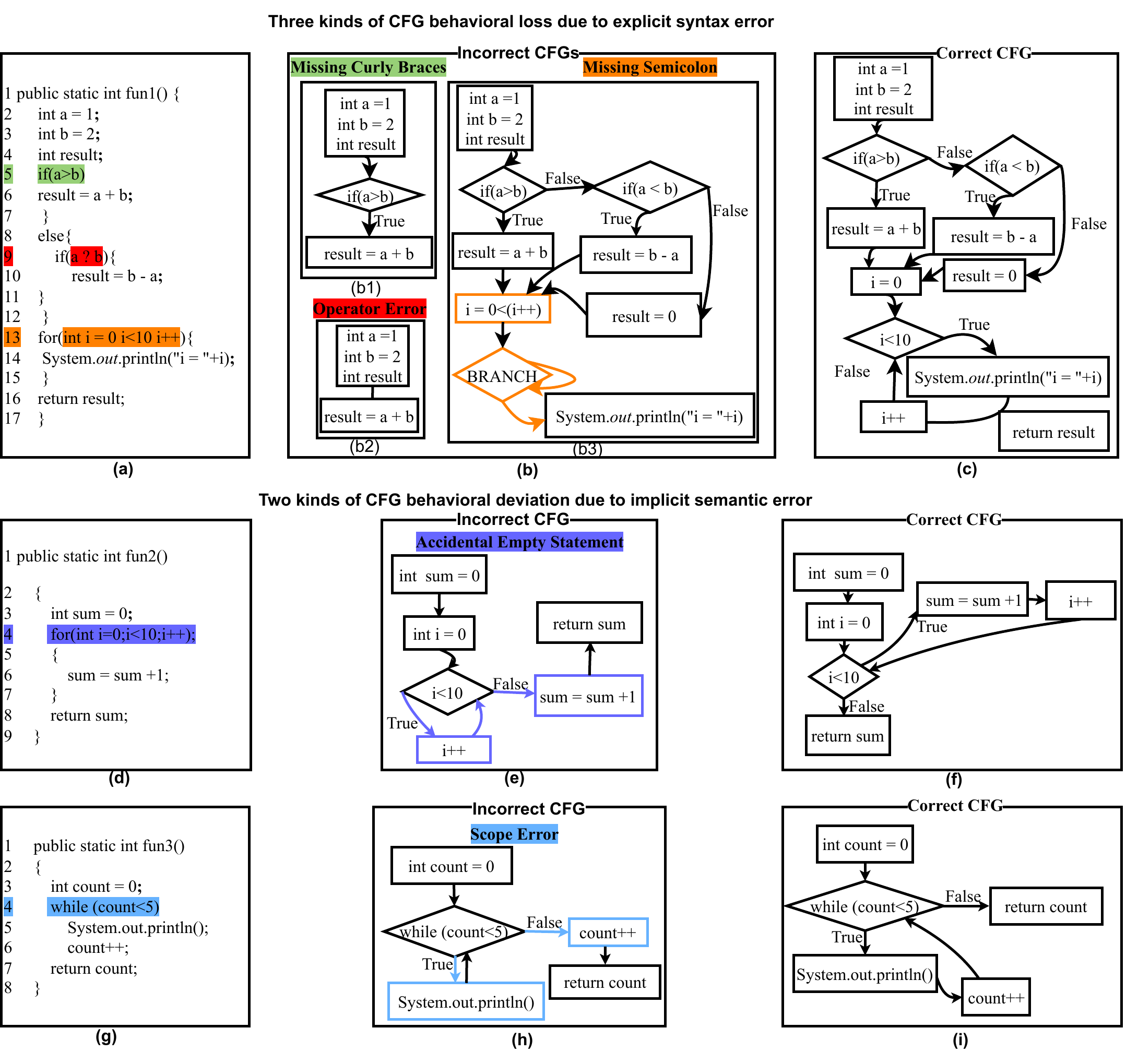}
     \vspace{-2mm}
    \caption{CFG Behavioral Loss or Deviation Caused by Explicit Syntax Errors or Implicit Semantic Errors in Java Code}
    \label{fig: killing example}
    \vspace{-3mm}
\end{figure*}

Fig.~\ref{fig: killing example}-a shows Java code with three syntax errors, all causing behavioral loss in the generated CFG. The missing curly brace (green part) causes the AST-based method to misinterpret the if statement's closing bracket as a method's closing bracket, leading to behavioral loss in the generated CFG (Fig.~\ref{fig: killing example}-b1). The operator error (red part) results in behavioral loss in the generated CFG because the AST-based method cannot traverse the entire AST structure (Fig.~\ref{fig: killing example}-b2). The absence of a semicolon (orange part) causes the for loop statement's expressions (i.e, \textit{int i=0}; \textit{i\textless10}; \textit{i++}) to be incorrectly treated as loop variable initialization statements, causing behavioral loss in the generated CFG (Fig.~\ref{fig: killing example}-b3).

Bad coding practices can result from carelessness or accidental violation of coding conventions, leading to code that compiles but has a different semantics than the developer originally intended.
These practices cause implicit semantic errors such as accidental empty statements or misleading indentation, leading to behavioral deviation in the generated CFG. 
For instance, Fig.~\ref{fig: killing example}-d shows an accidental empty statement (purple part) caused by a misplaced semicolon at the end of a for loop. 
The loop executes 10 times without performing any action (Fig.~\ref{fig: killing example}-e). The intended behavior is to increment the \textit{sum} variable by 1 within each loop iteration, but the generated CFG deviates from the expected behavior. 
Similarly, Fig.~\ref{fig: killing example}-g illustrates a scope error (blue part) caused by the lack of curly braces around the intended loop body, leading to an infinite loop and the \textit{count} variable never being incremented (Fig.~\ref{fig: killing example}-h), again resulting in behavioral deviation in the generated CFG.

\begin{figure}[t]
    \centering
    \includegraphics[width=0.45\textwidth]{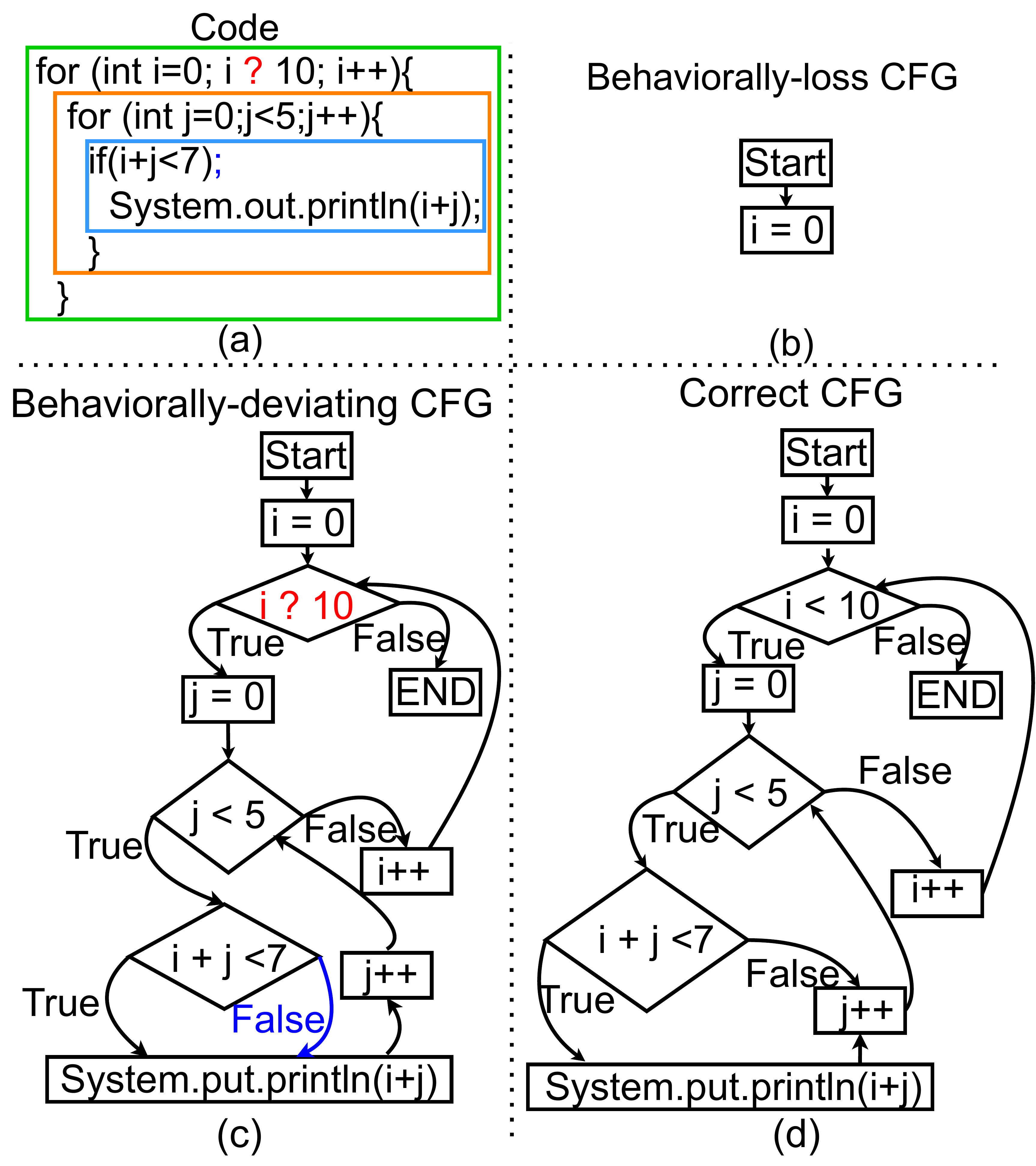}
    \caption{Motivation of Hierarchical Task Breakdown}
    \label{fig:Hierarchical Task Breakdown Example}
    \vspace{-6.5mm}
\end{figure}

\begin{figure*}[t]
    \centering
    \includegraphics[width=1.0\textwidth]{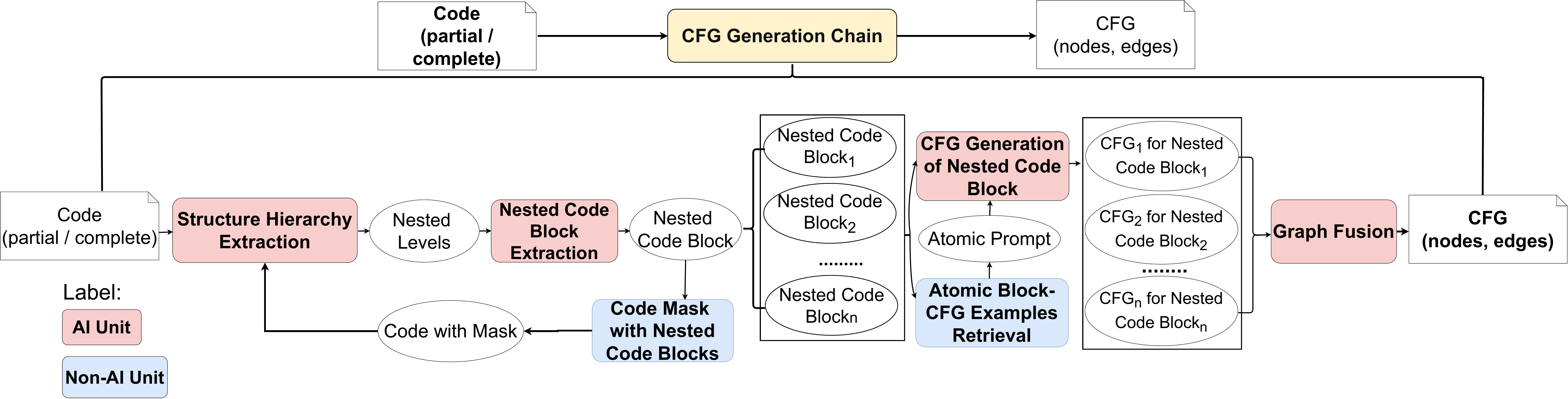}
    \caption{Overall Framework of CFG-Chain}
    \label{fig: Overall Framework of Our Approach}
    \vspace{-3.5mm}
\end{figure*}

To address the issue of behavioral loss and deviation of CFG due to syntax errors and bad coding practices, we treat source code as a natural language and use pre-trained large language models (LLMs, such as GPT-3.5~\cite{openai_GPT-3.5} and CodeX~\cite{chen2021evaluating}) to understand it~\cite{Devanbu2012OnTN, Allamanis2018ASO}. LLMs are robust in processing natural language, capable of handling common grammatical errors and semantic errors (e.g., misspelled words) while understanding sentence meaning accurately. This robustness stems from pre-training on vast amounts of text data, allowing LLMs to learn a wide range of language patterns and contextual information~\cite{wan2022they, karmakar2021pre, jiang2021treebert, wang2021codet5, poesia2022synchromesh}. For example, LLMs can tolerate grammatical errors and misspelled words, such as ``They is going to the park.'' and ``He is a engineer.'', based on context and common sense to understand the intended meaning. With this robustness, LLMs can prevent behavioral loss even in the presence of explicit syntax errors in code. Moreover, LLMs can detect implicit semantic errors based on context, avoiding behavior deviation. For instance, in Fig.~\ref{fig: killing example}-g, LLMs can infer that lines 5 and 6 belong to a while loop by analyzing the count variable in lines 3, 4, and 6 and the indentation in lines 5 and 6, even without curly braces.

When using LLMs, there are two primary approaches: supervised fine-tuning~\cite{liu2023pre, Huang2023APIEA, schick2020automatically} and unsupervised in-context learning~\cite{bommasani2021opportunities, Brown2020LanguageMA, raffel2020exploring}.
Supervised fine-tuning requires labeled training data, whereas in-context learning does not. 
Recent studies have shown that in-context learning is effective for code-related tasks~\cite{huang2022se, shi2022xricl, tenney2019learn}. 
Due to its convenience and cost-effectiveness, we prioritize the use of in-context learning to generate CFG.
We exemplify and evaluate our approach on Java code, but our approach does not make any assumptions of specific language syntax or features and thus can be applied to other statically-typed program languages.

Generating CFG nodes and edges for Java code directly using LLMs is challenging due to their uncertainty, errors, and hallucination problems ~\cite{ji2023survey, creswell2022faithful, yao2022react}. For instance, two code statements ``\textit{if(i==1) {return true}}'' may be treated as one node, instead of being treated as two separate nodes. 
% CoT
To mitigate this problem, we design an informative Chain of Thought (CoT)~\cite{wang2022self, wei2022chain} to CFG generation, which involves four steps: \textit{structure hierarchy extraction} to identify nested levels, \textit{nested code block extraction} to obtain code blocks at each level, \textit{CFG generation of nested code blocks}, and \textit{graph fusion} to integrate all nested code blocks' CFGs.

However, the original CoT method has limitations due to its use of a single prompt to implement all the step responsibilities, which can lead to error accumulation and the creation of an ``epic'' prompt with too many step duties that are difficult to optimize and control. 
To overcome these limitations, we adopt the principle of single responsibility in software engineering and break down the CoT into an AI chain~\cite{wu2022promptchainer,wu2022ai}, with each step corresponding to a se AI-unit.
We develop an effective prompt for each AI-unit which performs separate LLM calls. 
This AI chain can interact with LLMs step by step to generate CFG for source code, regardless of whether the Java code is fully compilable or partially uncompilable.

We conduct several experiments to evaluate the performance of our CFG generation approach and compare it with existing methods~\cite{Soot,pawlak2016spoon}.
Our results show that our approach has a strong error-tolerance ability in generating CFGs for code with explicit syntax errors, with a higher node coverage by 35\% and 95\% compared to the AST-based method~\cite{pawlak2016spoon}. 
Moreover, our approach demonstrates a strong understanding ability in generating CFGs for code with implicit semantic errors, with a higher edge coverage by 9.6\% compared to the AST-based method~\cite{pawlak2016spoon} and 14\% compared to the bytecode-based method~\cite{Soot}.
We also conduct an ablation experiment to investigate why our AI Chain performed well, which shows that our AI chain design was reasonable.
Finally, we summarize our findings as three AI chain design principles:
Hierarchical Task Breakdown, Unit Composition, and Mix of AI Units and Non-AI Units.
These principles can serve as guidelines for future prompt engineering projects in software engineering.

The main contributions of this paper are as follows:
\begin{itemize}[leftmargin=*]
% \vspace{-1mm}
    \item
    We find the incomplete and inaccurate CFG generated by existing methods can be attributed to explicit syntax errors and implicit semantic errors resulting from poor coding practices.
    \item
    We propose a novel approach that leverages the error-tolerant and understanding ability of LLMs, treating code as a natural language, to generate CFGs.
    \item 
    To address the limitations of using an ``epic'' prompt, we break down the CoT into an AI chain with multiple AI units, based on the principle of single responsibility, which improves the robustness of LLM outputs.
    \item
    Our experimental results demonstrate the superiority of our approach over traditional methods, and its strong adaptability to various scenarios.
    \item
    We present a set of practical principles for employing prompt engineering in software engineering tasks.
\end{itemize}

\section{APPROACH}

Generating behaviorally-accurate CFGs for statically-typed partial code is challenging due to the common problems of behavioral loss and deviation. 
Behavioral loss occurs when a CFG loses some nodes, resulting in inaccurate program behavior, often due to explicit syntax errors when using the AST-based method.
Behavioral deviation refers to a CFG that deviates from expected behavior, making it difficult for developers to understand and debug the code, often due to implicit semantic errors caused by bad coding practices.

Our approach CFG-Chain, addresses these challenges by leveraging the contextual understanding capability of LLMs, which can tolerate code containing explicit syntax errors and detect implicit semantic errors. 
We simulate the human thought process, breaking down the task into single-responsibility sub-problems and designing functional units. 
These units are linked in a serial, parallel, or split-merge structure to create a multi-round interaction with the LLM to solve problems step by step. 
We use CodeX~\cite{chen2021evaluating} as our underlying LLM, and our approach focuses on what problem to solve, including task characteristics, data properties, and information flow, by standing on the shoulder of CodeX. 
This approach differs from fine-tuning LLMs, which requires significant effort in data gathering, preprocessing, annotation, and model training.

\subsection{Hierarchical Task Breakdown}

When faced with a piece of code containing an explicit syntax error (red question mark) and an implicit semantic error (blue semicolon) (Fig.~\ref{fig:Hierarchical Task Breakdown Example}-a), the bytecode-based method cannot generate a CFG, and the AST-based method can only produce a behaviorally-losing CFG (Fig.~\ref{fig:Hierarchical Task Breakdown Example}-b). 
However, a complete CFG can be generated using LLMs (e.g., CodeX). 
As shown in Fig.~\ref{fig:Hierarchical Task Breakdown Example}-c, even if there is an explicit syntax error (red question mark), the nodes are not lost, thanks to the LLM's ability to tolerate such errors. 
However, the generated CFG may still exhibit behavioral deviation due to implicit semantic errors caused by bad coding practices. 
For example, an accidental empty statement caused by a semicolon at the end of an '\textit{if}' condition can lead to incorrect program behavior. 
As the code's nested structure becomes more complex, it becomes increasingly difficult to detect implicit semantic errors using a single LLM call and a single instruction to ``generate the given code's CFG''. 
To address this issue, we need to make the instruction more informative and break it down into several sub-instructions, each executed by a separate LLM call, to detect implicit semantic errors and generate a behaviorally-correct CFG, as shown in Fig.~\ref{fig:Hierarchical Task Breakdown Example}-d.

To develop a reasonable decomposition, we analyzed the code in Fig.~\ref{fig:Hierarchical Task Breakdown Example}-a again, and found that the code has hierarchical nesting relationships with three layers identified by green, orange, and blue borders. 
Each layer contains code blocks that can be processed in the same way to generate the CFG for that layer. Hence, we devised a ``recursively nested code replacement'' method to process the nested code blocks layer by layer and generate the complete code CFG. 
The method begins with the innermost block and converts it to CFG, replaces it with the specified block string (referred to as ``code masking''), and works outward through the higher nesting levels until all nesting levels are replaced. 
This method enables us to break down the task of CFG generation into a chain of AI with many manageable units, as demonstrated in Fig.~\ref{fig: Overall Framework of Our Approach}.

The first AI unit, called \textit{Structure Hierarchy Extraction}, identifies the nested levels within a code structure, which guides the next AI unit, called \textit{Nested Code Block Extraction}, to extract the code blocks under each nested level. We use a non-AI unit called Code Mask with Nested Code Blocks to replace the extracted code blocks with the specified block string and update the code. The process repeats the execution of the two AI units until all nesting levels are processed.
The nested code blocks are then inputted to the AI unit called \textit{CFG Generation of Nested Code Block}, which generates individual CFGs for each block. The non-AI unit called \textit{Atomic Block-CFG Examples Retrieval} helps by offering prompt examples specific to the nested code block. Finally, the AI unit called \textit{Graph Fusion} integrates the nodes and edges of every nested code block’s CFG, resulting in a behaviorally-correct CFG.

\begin{figure}[t]
    \centering
    \includegraphics[width=0.45\textwidth]{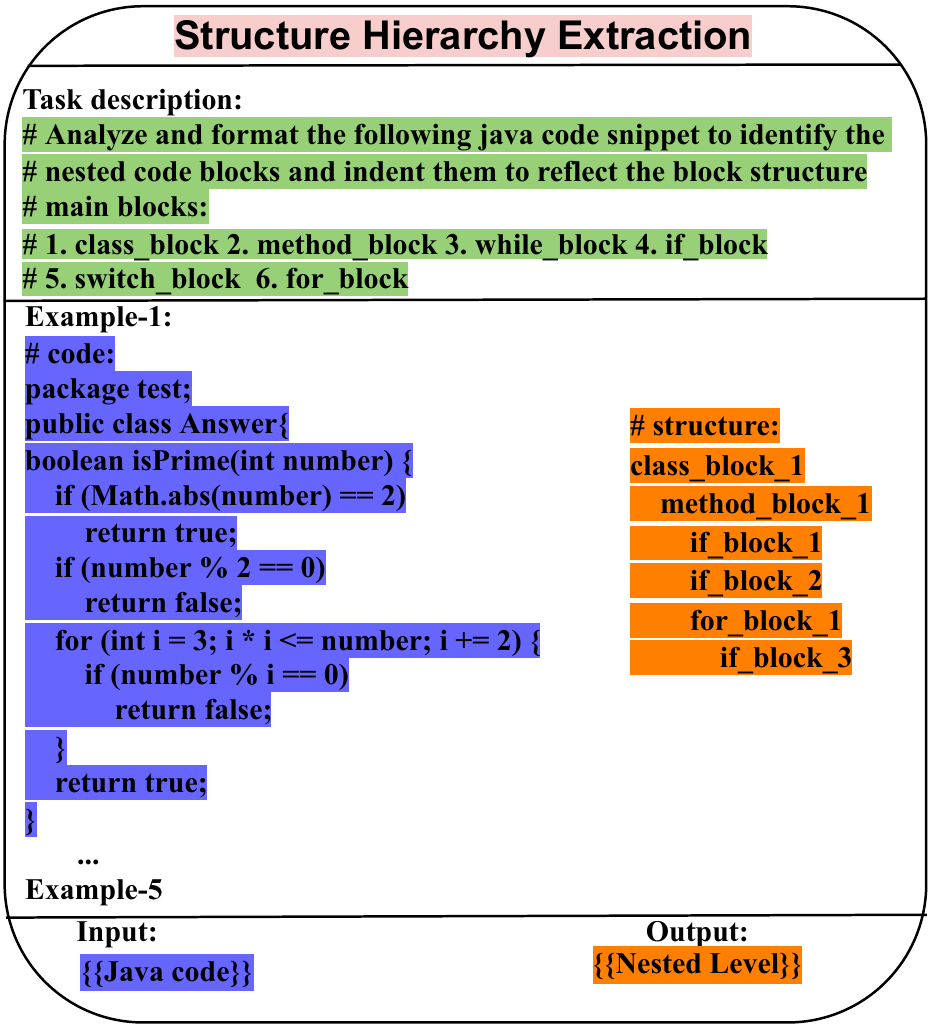}
    \caption{Structure Hierarchy Extraction Unit}
    % \vspace{-6mm}
    \label{fig:unit1}
\end{figure}

\begin{figure}[t]
    \centering
    \includegraphics[width=0.45\textwidth]{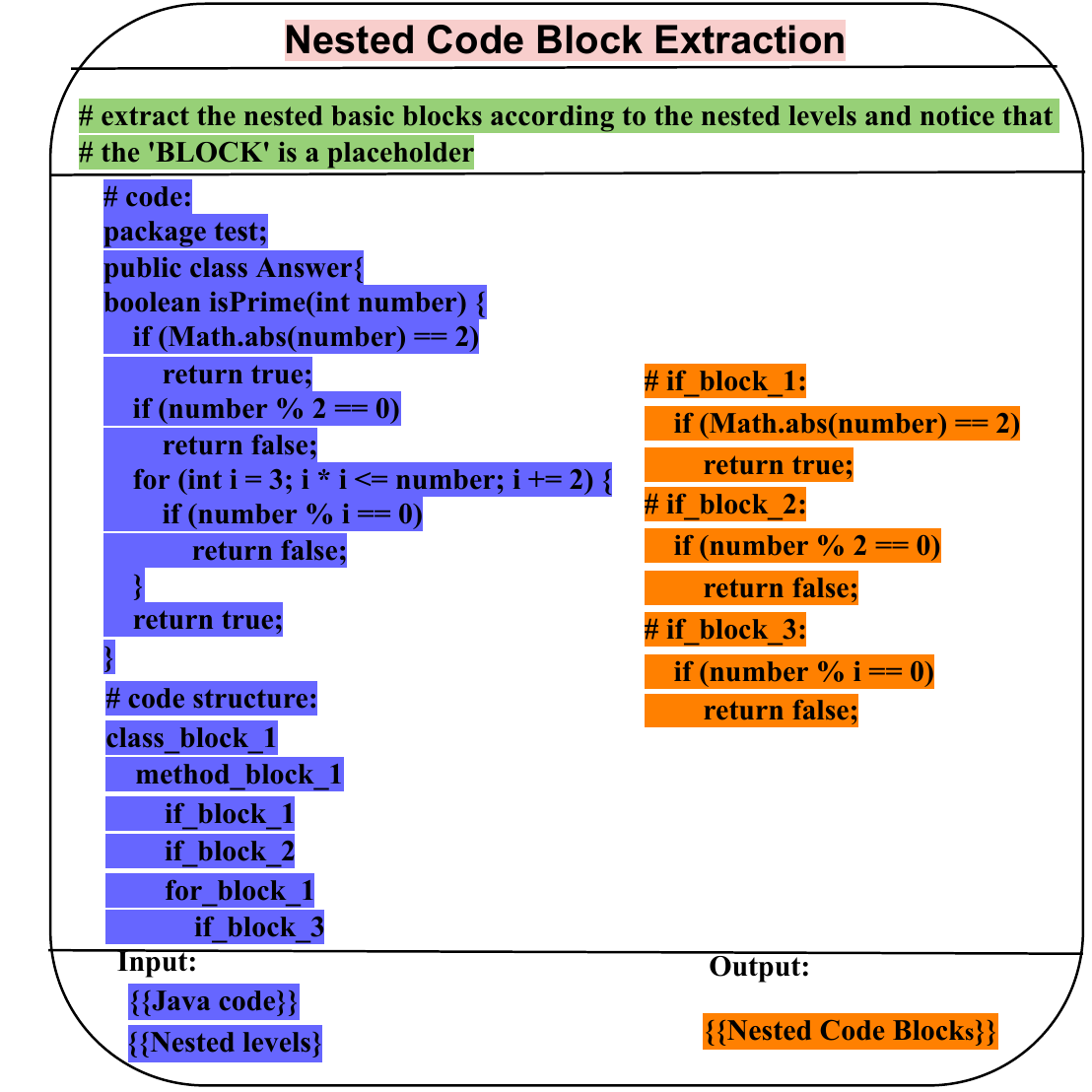}
    \caption{Nested Code Block Extraction Unit}
    \label{fig:unit2}
    \vspace{-3mm}
\end{figure}

% \vspace{-4mm}
\subsection{Prompt Design for AI-Units}
To generate an accurate CFG for statically-typed partial code, we break down the task into multiple units and leverage the strengths of both AI and non-AI units. This section focuses on describing how to write natural language prompts that program LLMs to perform various functionalities of AI units.

An empirical study~\cite{huang2022se} showed that task description and examples are critical for prompt design. To standardize our prompt design, we devised a generic template that includes a task description and a set of input-output examples. We describe the construction of the template using the \textit{Structure Hierarchy Extraction Unit} as an example, which extracts the structure hierarchy of the given code.
As shown in Fig.~\ref{fig:unit1}, at the top of the template is a description (e.g., ``Analyze and format the following...'') in green, in the middle are five input-output examples (e.g.,  Input: ``code: package test: public class Answer...'', Output: ``structure: class\_block\_1...''), and below are an input (e.g., a java code) and an output (e.g., the extracted structure hierarchy of the given code).
To save space, we illustrate input and output side by side.
But the input and output are sequentially consecutive in the real prompt.

Noted that in this work, we pre-select five examples that are used for all AI units. While the model adaptability generally increases with more examples~\cite{huang2022se}, Min et al.~\cite{min-etal-2022-rethinking} have shown that additional examples beyond four results in limited increase in accuracy.
In addition, when selecting the five examples, we also consider their representativeness and diversity.
For example, for the \textit{Structure Hierarchy Extraction Unit}, different examples should have different syntax error (e.g., missing curly braces, missing semicolon, etc.), semantic error (e.g., accidental empty statement) and code structure (e.g., ``if'', ``while'', ``for'' structures).

In the following sections, we describe the prompt design of each of the four units.

\subsubsection{Structure Hierarchy Extraction Unit}
This AI unit is responsible for extracting the nested levels of the given code.
To prompt the LLM to perform this task, a generic template is used, as shown in Fig.~\ref{fig:unit1}, with a task description of ``Analyze and format the following java...'', five examples, and a space to input the code to be processed to obtain its structure.
To improve recognition at the nesting level, the task description also includes six common types of code blocks, such as class declarations, method declarations, and if statements.

\begin{figure}[t]
    \centering
    \includegraphics[width=0.40\textwidth]{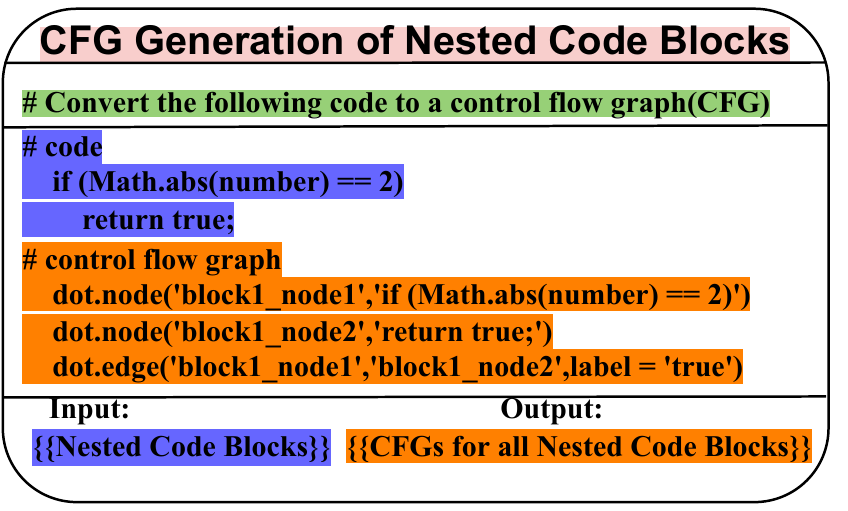}
    \caption{Nested Code CFG Generation Unit}
    \label{fig: unit3}
    % \vspace{-6mm}
\end{figure}

\begin{figure}[t]
    \centering
    \includegraphics[width=0.48\textwidth]{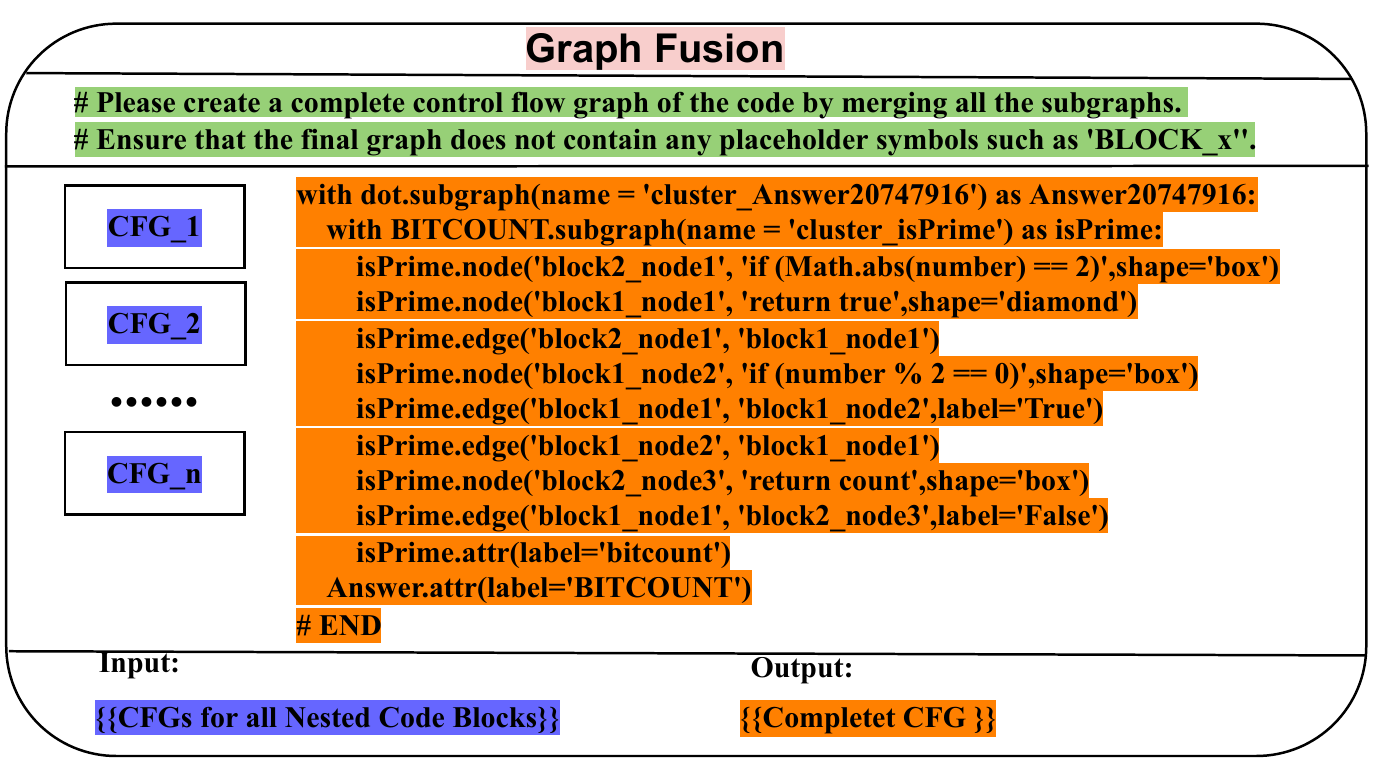}
    \caption{Graph Fusion Unit}
    \label{fig:unit4}
    \vspace{-4mm}
\end{figure}

\subsubsection{Nested Code Block Extraction Unit}
This AI unit is responsible for extracting the basic blocks according to the code structure. Fig.~\ref{fig:unit2} illustrates the contents of the prompt, which includes the task description ``Extract the nested code block according to code structure...'' along with five corresponding examples. 
Each example consists of two inputs, the code and its corresponding code structure, and their respective outputs, the nested code block.
Note that this unit prioritizes the processing of non-nested code structures, as shown in Fig.~\ref{fig:unit2}. 
In the given example, the three ``if'' blocks are non-nested, while the ``for'' block contains an ``if'' block (if\_block\_3), so the if\_block\_3) block is processed before the ``for'' block. Further details are provided in Section~\ref{running example}.

\subsubsection{Nested Code CFG Generation Unit}
This AI unit is designed to generate the nodes and edges of all nested code blocks’ CFG. 
Fig.~\ref{fig: unit3} shows the prompt content of this unit, which includes a task description, ``Convert the following code to a control flow graph (CFG),'' and five examples.
Each example in prompt includes a basic block input and the corresponding CFG output.
These examples train the model to mimic the behavior characteristics of the CFGs.
When a code block is inputted to the unit, it outputs the corresponding CFG.
To provide more effective prompts, we adopt a simple example retrieval strategy. 
Specifically, we prepare five examples for each of the six types of nested code blocks (i.e., class\_block, method\_block, while\_block, if\_block, switch\_block, and for\_block). 
These examples constitute our knowledge base. 
Then, given a nested code block type, we use five examples of the same type in prompt that match from our existing knowledge base.
And each type of nested code block is representative and diverse.
e.g., for while\_block, the example ``while ( if(...) )...'' or ``while ( for ( )'' contains for\_block and if\_block.
Note that this unit can only process one basic block at a time, and if a code has multiple basic blocks, we execute multiple basic block generation units in parallel as illustrated in Figure~\ref{fig: Overall Framework of Our Approach}.
In addition, we use Graphviz~\footnote{\href{https://graphviz.org/}{\textcolor{blue}{https://graphviz.org/}}} to visualize the final CFG. Graphviz requires the CFG to be in Python code format, so we use LLMs to generate a Python-like code for the CFG.

\subsubsection{Graph Fusion Unit}
We design this unit to integrate the nodes and edges of the respective nested code blocks' CFGs, resulting in a comprehensive CFG for the given code. 
The prompt for this unit is shown in Fig.~\ref{fig:unit4}, which includes a task description ``Please create a complete control flow graph of the code...'', and five examples. 
The input for this unit is CFGs for nested code blocks. 
The output is the complete CFG formed by fusing the multiple CFGs. 
By providing the CFGs for nested code blocks to this unit, it can learn to mimic the behavior characteristics of the given examples and produce a complete and accurate CFG for the given code.

\begin{figure*}[h]
    \centering
    \includegraphics[width=0.99\textwidth]{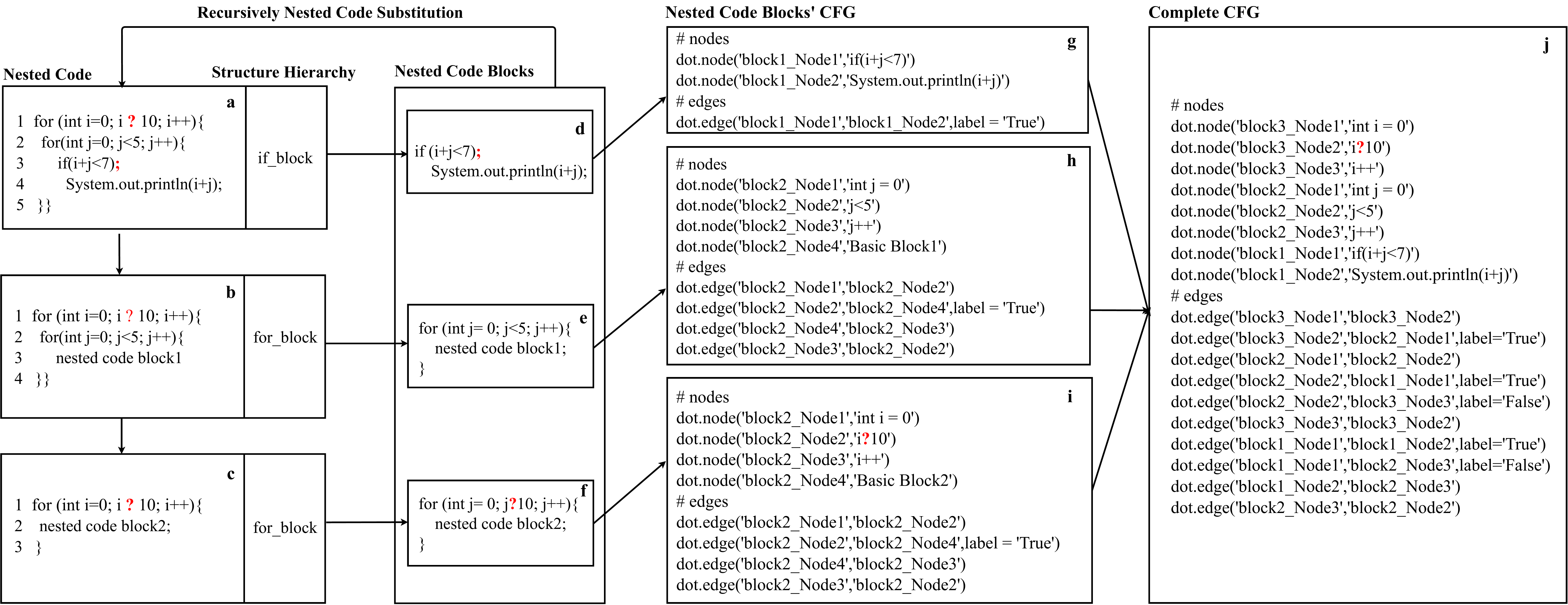}
    \caption{Running Example}
    \label{fig:Running Example}
\end{figure*}

\subsection{Running Example}\label{running example}
To demonstrate how the AI units work together and how the data is transformed among these AI units, we present an example using a nested java code with an operator error and an unexpected null statement, illustrated in Fig.~\ref{fig:Running Example}-a.

The first step is to input the code into the \textit{Structure Hierarchy Extraction Unit}, which identifies the innermost nesting structure, in this case, an if\_block.
If multiple nested structures exist side-by-side in the code, the \textit{Structure Hierarchy Extraction Unit} extracts them simultaneously. 
For example, given the code ``for(...)\{if\{...\}...if\{...\}\}'', this unit outputs if\_block1 and if\_block2.

Next, the nested code and if\_block are input to the \textit{Nested Code Block Extraction Unit}, which extracts the nested code block from the nested code based on if\_block. 
The output of this unit is shown as Fig.\ref{fig:Running Example}-d. 
Then, we use the extracted nested code block to replace the corresponding part of the original nested code and obtain the masked nested code, as shown in Fig.\ref{fig:Running Example}-b. 
Subsequently, the masked nested code is input to the \textit{basic block extraction unit} to obtain the innermost nesting structure, i.e., for\_block. 
We repeat these steps until we get the outermost nested code block, as shown in Fig.~\ref{fig:Running Example}-f.

Once we have all the nested code blocks, we input them into the \textit{CFG Generation of Nested Code Block Unit} to generate their CFGs.
In this example, the nested code block if (Fig.~\ref{fig:Running Example}-d), for (Fig.~\ref{fig:Running Example}-e), and for (Fig.~\ref{fig:Running Example}-f) are converted to their respective CFG-1 (Fig.~\ref{fig:Running Example}-g), CFG-2 (Fig.~\ref{fig:Running Example}-h), and CFG-3 (Fig.~\ref{fig:Running Example}-i).
Note that this process is executed in parallel, using multiple \textit{CFG Generation of Nested Code Block Units}.

Finally, we input the CFGs into the \textit{Graph Fusion} Unit to generate a complete CFG for the code, as shown in Fig.~\ref{fig:Running Example}-j.

\section{EXPERIMENTAL SETTING}
In this section, we present our research questions to evaluate the performance of our approach, along with our experimental setup. This includes data preparation, baselines, and evaluation metrics.

\subsection{Research Question}
We formulated three research questions to assess the performance of CFG-Chain in generating CFGs:
\begin{itemize}
    \item RQ1: The quality of each AI unit.
    \item RQ2: The performance of CFG-Chain in CFG generation.
    \item RQ3: The ablation study of CFG-Chain.
\end{itemize}

\subsection{Data Preparation}
To evaluate our approach, we collect 90,000 error-free, compilable code samples from a reliable reference~\cite{singh2013elements}. However, since these code samples do not contain explicit syntax or implicit semantic errors, we randomly divide them into three groups, each containing 30,000 samples, and select subsets of 240 code samples from each group. We ensure that each sample has at least two levels of nesting.

The first subset is the NC dataset, containing error-free, compilable code samples.
For the second subset, we manually introduce missing curly braces, missing semicolons, and missing operator errors separately into three groups of 80 code samples each (see Fig.~\ref{fig: killing example}-a). This subset is the ESE dataset, containing code samples with explicit syntax errors.
For the third subset, we manually introduce missing curly braces and missing operator errors separately into two groups of 120 code samples and 120 code samples, respectively (see Fig.~\ref{fig: killing example}-d and Fig.~\ref{fig: killing example}-g). This subset is the ISE dataset, containing code samples with implicit semantic errors.

In summary, we prepare three distinct datasets:

\begin{itemize}
    \item  NC dataset (with 240 error-free, compilable code samples)
    \item ESE dataset (with 80 samples containing missing curly braces, 80 with missing semicolons, and 80 with missing operator errors)
    \item ISE dataset (with 120 samples containing accidental empty statements, and 120 with scope errors)
\end{itemize}

% the NC dataset (with 240 error-free, compilable code samples), the ESE dataset (with 80 code samples containing missing curly braces, 80 with missing semicolons, and 80 with missing operator errors), and the ISE dataset (with 120 samples containing accidental empty statements, and 120 with scope errors).

\begin{figure}[t]
    \centering
    \includegraphics[width=0.5\textwidth]{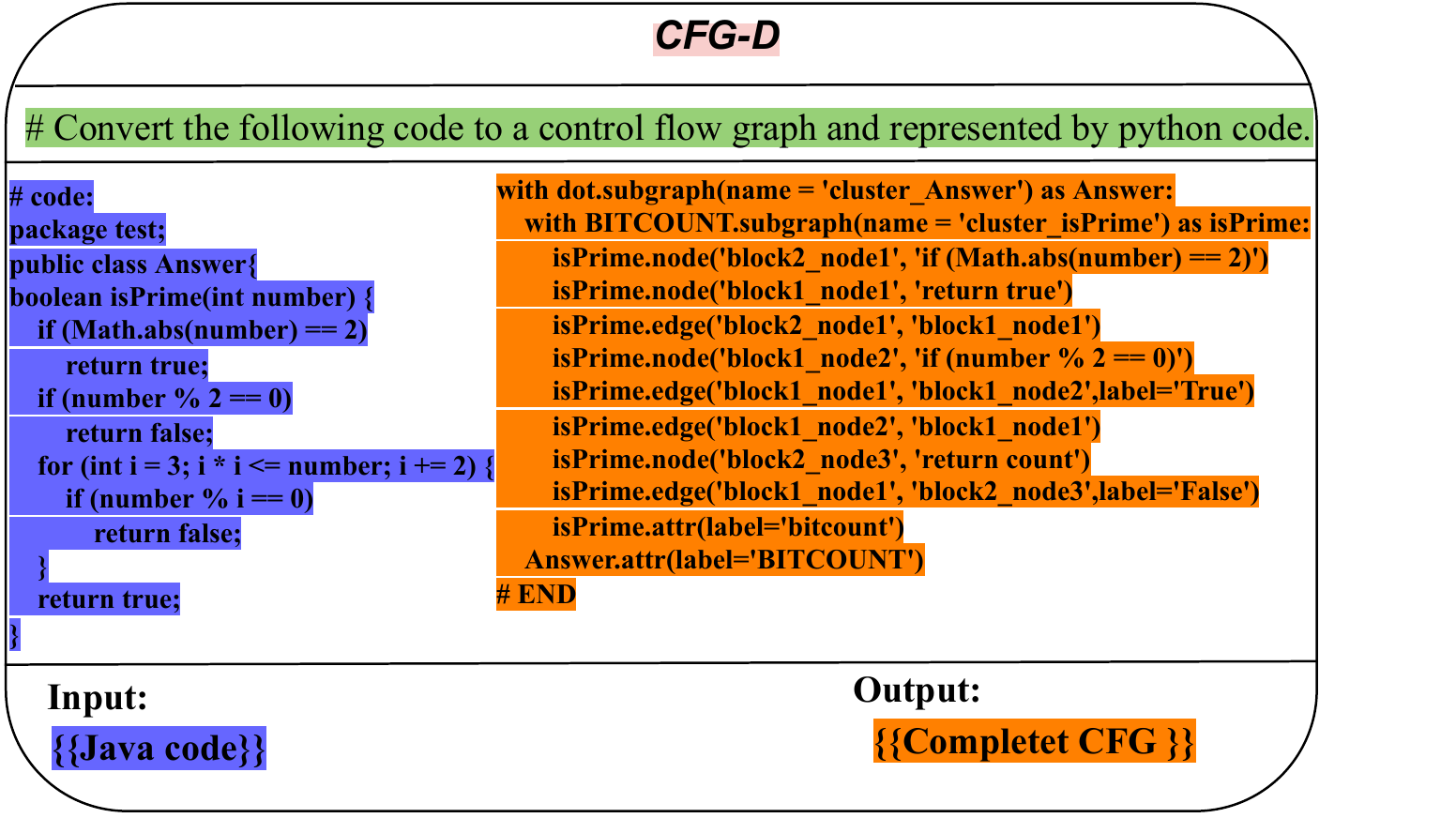}
    % \vspace{-4mm}
    \caption{Consult LLM directly (CFG-D)}
    \label{fig:TAF}
     \vspace{-5mm}
\end{figure}

\begin{figure}[t]
    \centering
    \includegraphics[width=0.5\textwidth]{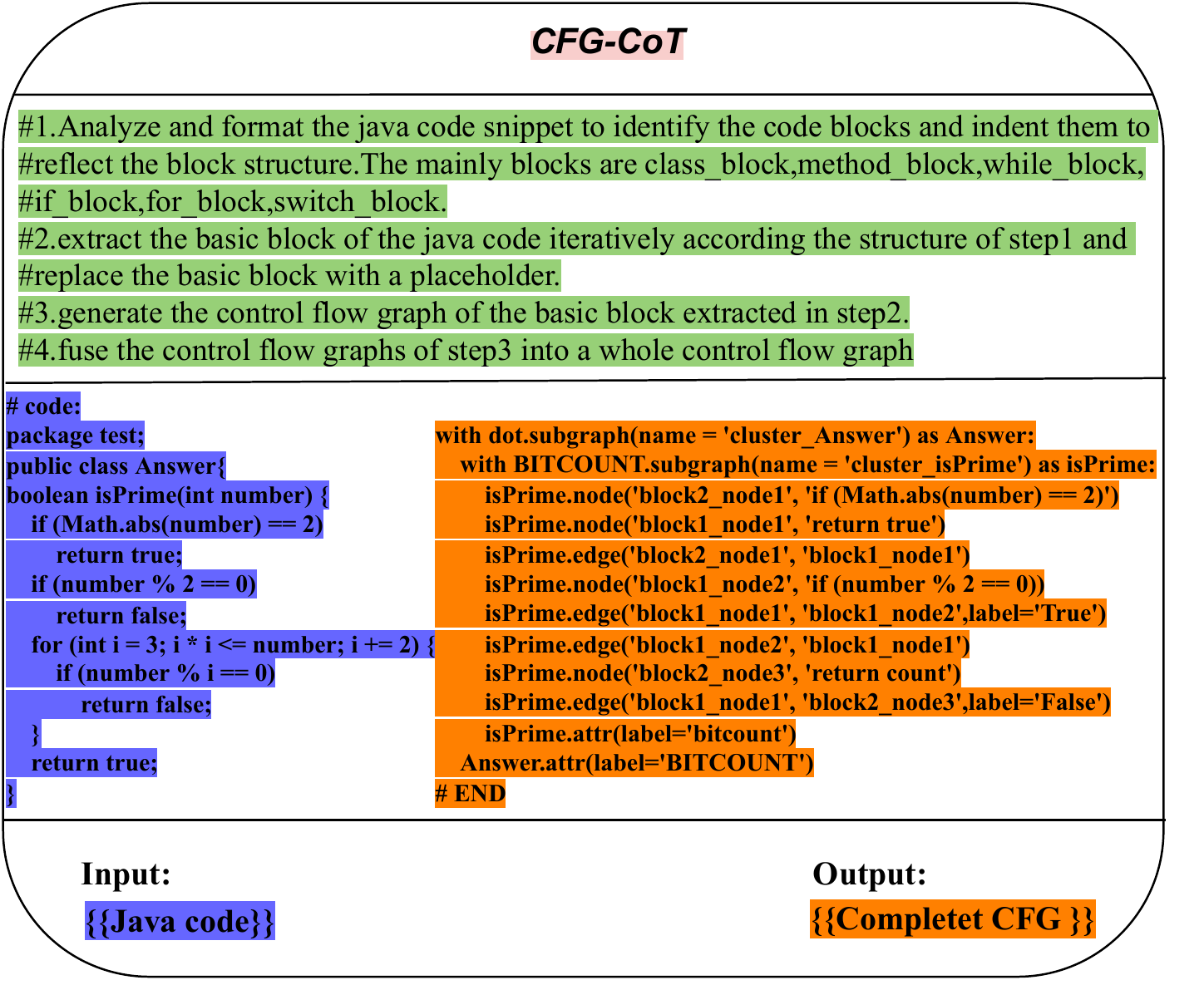}
    % \vspace{-4mm}
    \caption{Consult LLM based on CoT (CFG-CoT)}
    \label{fig:CoT}
    \vspace{-3mm}
\end{figure}

\subsection{Baselines}
To evaluate the effectiveness of CFG-Chain in CFG generation, we compare it with existing methods for CFG generation, which fall into two categories: bytecode-based methods like Soot~\cite{Soot}, and AST-based methods like Spoon~\cite{pawlak2016spoon}. We run CFG-Chain, Soot\footnote{\href{https://github.com/soot-oss/soot}{\textcolor{blue}{https://github.com/soot-oss/soot}}}, and Spoon\footnote{\href{https://github.com/INRIA/spoon}{\textcolor{blue}{https://github.com/INRIA/spoon}}} on the NC, ESE, and ISE datasets to generate CFGs, and then compare their performance.

In addition, we conduct an ablation study of CFG-Chain to explain why it works. We design three variants for this purpose:
\begin{itemize}[leftmargin=*]
    \item
    CFG-D (see Fig.~\ref{fig:TAF}), which directly calls the LLM to generate the CFG of the Java code.
    \item
    CFG-CoT (see Fig.~\ref{fig:CoT}), a single-prompting approach that describes all steps in one chunk of prompt text and completes a single generative pass.
    \item 
    CFG-$Chain_{w/oAPR}$, a multiple-prompting approach that does not dynamically retrieve related examples with similar structures to specific atomic blocks.
\end{itemize}
To evaluate the effectiveness of CoT design, we compare CFG-D to CFG-CoT. To verify the effectiveness of AI chain design, we compare CFG-CoT to CFG-Chain. Finally, we evaluate the effectiveness of the atomic block-CFG examples retrieval strategy by comparing CFG-$Chain_{w/oAPR}$ to CFG-Chain.

\subsection{Evaluation Metrics}
\label{metrics}
In RQ1, we use accuracy as the evaluation metric for three tasks: \textit{structure hierarchy extraction}, \textit{nested code block extraction}, and \textit{CFG generation of nested code blocks}.
Accuracy is a binary metric that indicates whether the output of each unit is correct or not, with a value of 1 indicating correct output and 0 indicating incorrect output.
For \textit{graph fusion}, we use node coverage and edge coverage as the evaluation metrics. 
This is because the CFG generated by this unit is expected to reduce the number of nodes and edges due to explicit syntax errors and implicit semantic errors.
In RQ2 and RQ3, we use node coverage and edge coverage to evaluate the CFG generated from the Java code. These metrics provide a measure of how accurately the CFG captures the behavior of the code.

To calculate node coverage and edge coverage, we follow a specific procedure.
We enlist 12 master's students, each with over three years of Java development experience, to act as annotators for drawing CFGs. 
Each group of 4 students is assigned to one of the three datasets (NC, ESE, and ISE), and each student draws CFGs for 60 code samples within their assigned dataset. 
All the CFGs drawn by the students are used as standard answers for evaluation purposes.
After generating CFGs using the baselines and our approach, the resulting CFGs are compared with the standard answers by annotators to determine their correctness. Three factors are considered when evaluating the correctness of the resulting CFGs:

\begin{itemize}[leftmargin=*]
    \item Number of nodes and edges: Two CFGs should have the same number of nodes and edges. If the resulting CFG is missing a node or edge, it is deemed incorrect.
    \item Node and edge labels: Node and edge labels should accurately represent the program elements they represent. For instance, a node should represent a statement or control structure, while an edge should represent process control. If the label is incorrect, the resulting CFG is considered incorrect.
    \item Correctness of process control: Process control must accurately reflect the program's control process. For instance, conditional branching should choose the proper branch based on the outcome of the conditional statement. An incorrect flow control results in an incorrect Control Flow Graph (CFG).
\end{itemize}

To determine the correctness of the resulting CFGs, the number of nodes and edges, the accuracy of the labels, and the correctness of the flow control are compared.
If the resulting CFG is incorrect, we identify the nodes and edges in the standard answer that caused the incorrect results and count their number as "wrong". Then, we calculate the total number of nodes and edges in the standard answer as "total". We then calculate the node coverage as (total-wrong)/total, and edge coverage as (total-wrong)/total, respectively. The higher the node coverage and edge coverage, the better the method of generating CFG.

% \vspace{1mm}
\section{EXPERIMENTAL RESULTS}
This section delves three research question to evaluate and discuss the performance of our approach.
% \vspace{2mm}
\subsection{RQ1: The quality of each AI unit}
\vspace{2mm}
\subsubsection{Motivation}
The CoT approach inspires us to break down complex tasks into simple steps. 
However, the use of a single "epic" prompt in CoT-based methods limits its effectiveness and can lead to error accumulation. 
To address this, we develop an AI chain with explicit sub-steps, where each step corresponds to a separate AI unit.
In this RQ, we investigate whether each AI unit in our approach can effectively ensure the accuracy of CFG generation.

\subsubsection{Methodology}
We apply CFG-Chain to the NC, ESE, and ISE datasets and collect the intermediate results produced by each AI unit.
These intermediate results are then provided to master students to calculate the metric values, such as accuracy, node coverage, and edge coverage. 
The results are presented in Table~\ref{table: The performance of each AI unit}, and more detailed information can be found in Section~\ref{metrics}.

\subsubsection{Result Analysis}
Table \ref{table: The performance of each AI unit} presents the experimental results of running CFG-Chain on the NC, ESE, and ISE datasets. The first unit, \textit{structure hierarchy extraction}, demonstrates consistent performance across all three datasets, achieving an accuracy of 0.82 on NC, 0.80 on ESE, and 0.82 on ISE. This indicates that the unit can effectively extract the structure hierarchy, and our approach has strong error-tolerant and understanding ability even when facing with syntax and semantic errors.

The second unit, \textit{Nested Code Block Extraction}, achieves an accuracy of 0.84 on NC, and 0.80 on both ESE and ISE datasets, suggesting that this unit is capable of accurately extracting nested code blocks.
In the third unit, \textit{Nested Code Block Generation}, we observe a higher accuracy of 0.89 on the NC dataset. However, its accuracy slightly decreases to 0.82 on ESE and 0.86 on ISE, indicating that the presence of explicit and implicit errors in the code may affect the performance of the CFG generation.

Regarding the fourth unit, \textit{Graph Fusion}, we observe strong node coverage across all datasets, with a value of 0.93 on both NC and ISE. However, edge coverage consistently scored lower, with values of 0.82 on NC and 0.80 on both ESE and ISE datasets. This is because the loss of a node results in the loss of all edges connected to that node, while the loss of an edge does not affect the node.

\vspace{0.66mm}
\noindent\fbox{\begin{minipage}{8.2cm} \textit{The high accuracy of AI units confirms that our prompt design is effective and conforms that our prompt composition is effective for connecting AI units to accomplish higher-layer tasks effectively.}\end{minipage}}

\begin{table}[t]
\centering
\caption{The performance of each AI unit}
\vspace{-3mm}
\label{table: The performance of each AI unit}
\begin{tabular}{|c|c|c|c|c|}
\hline
AI unit                                                                                      & Dataset & Acc   & \begin{tabular}[c]{@{}c@{}}Node \\ Coverage\end{tabular} & \begin{tabular}[c]{@{}c@{}}Edge \\ Coverage\end{tabular} \\ \hline
\multirow{3}{*}{\begin{tabular}[c]{@{}c@{}}Structure \\ Hierarchy\\ Extraction\end{tabular}} & NC      & 0.82  & -                                                        & -                                                        \\ \cline{2-5} 
                                                                                             & ESE     & 0.80  & -                                                        & -                                                        \\ \cline{2-5} 
                                                                                             & ISE     & 0.82  & -                                                        & -                                                        \\ \hline
\multirow{3}{*}{\begin{tabular}[c]{@{}c@{}}Nested \\ Code Block\\ Extraction\end{tabular}}   & NC      & 0.84 & -                                                        & -                                                        \\ \cline{2-5} 
                                                                                             & ESE     & 0.80  & -                                                        & -                                                        \\ \cline{2-5} 
                                                                                             & ISE     & 0.80  & -                                                        & -                                                        \\ \hline
\multirow{3}{*}{\begin{tabular}[c]{@{}c@{}}Nested\\ Code Block\\ Generation\end{tabular}}    & NC      & 0.89  & -                                                        & -                                                        \\ \cline{2-5} 
                                                                                             & ESE     & 0.82  & -                                                        & -                                                        \\ \cline{2-5} 
                                                                                             & ISE     & 0.86  & -                                                        & -                                                        \\ \hline
\multirow{3}{*}{Graph Fusion}                                                                & NC      & -     & 0.93                                                     & 0.82                                                     \\ \cline{2-5} 
                                                                                             & ESE     & -     & 0.87                                                     & 0.80                                                     \\ \cline{2-5} 
                                                                                             & ISE     & -     & 0.93                                                     & 0.80                                                     \\ \hline
\end{tabular}
% \vspace{-mm}
\end{table}

\subsection{RQ2:The performance of CFG-Chain in CFG generation.}
\vspace{2mm}
\subsubsection{Motivation}
We aim to compare our CFG generation approach
with Soot~\cite{Soot} and Spoon~\cite{pawlak2016spoon}, which are the leading CFG generation tools based on bytecode and AST, respectively.
We aim to investigate if our approach can outperform the existing methods in generating accurate and complete CFGs, especially in the presence of explicit syntax errors that can cause behavioral losses and implicit semantic errors that can cause behavioral deviation.
\subsubsection{Methodology}
We apply three CFG generation approachs (CFG-Chain, Soot and Spoon) to the NC, ESE and ISE datasets and calculate three metric values (i.e., accuracy, node coverage and edge coverage).
The results are presented in Table~\ref{table: RQ2 result} and more detailed information can be found in Section~\ref{metrics}.

\subsubsection{Result Analysis}
Table~\ref{table: RQ2 result} shows the results.
For the NC dataset, both the AST-based and bytecode-based CFG generation methods achieve perfect node and edge coverage (i.e., 1) since the code samples are complete and compilable. While the node and edge coverage of CFGs generated by CFG-Chain on this dataset is not as high as those generated by traditional methods, it still demonstrates a competitive performance.

The presence of syntax errors in the code samples of the ESE dataset poses a challenge for both traditional CFG generation methods. The AST-based method shows a significant drop in both node and edge coverage to 0.64 and 0.41, respectively, indicating a substantial loss of behavior in the generated CFGs. This is demonstrated by Fig.~\ref{fig: killing example}(b), which shows that three syntax errors in the code cause behavior loss in the generated CFGs. The bytecode-based method shows even worse performance, with node and edge coverage of 0, as it requires compilable bytecode files. However, CFG-Chain demonstrates a strong error-tolerance ability in generating CFGs for code with explicit syntax errors. The node and edge coverage of CFGs generated by CFG-Chain are significantly higher than those generated by the AST-based method, indicating that CFG-Chain can generate CFGs that suffer from much less behavior loss.

In the ISE dataset, the presence of semantic errors due to bad coding practices poses a challenge for both traditional CFG generation methods. Their node coverages of the CFGs are 1, but their edge coverage are low, at 0.73 and 0.70, respectively. This is because both methods can only generate a CFG based on the original behavior of the code, without considering whether there is a behavior deviation due to bad coding practices. This is demonstrated in two examples in Fig.~\ref{fig: killing example}(d) and (g).
In contrast, CFG-Chain demonstrates a strong understanding ability in generating CFGs for code with implicit semantic errors. The edge coverage of the CFGs generated by CFG-Chain is significantly higher than those generated by the traditional methods, indicating that CFG-Chain can intelligently avoid behavioral deviations.
\vspace{1mm}
\noindent\fbox{
\begin{minipage}{8.6cm} \textit{Standing on the shoulder of LLM for CFG generation, CFG-Chain is not limited by the explicit syntax error or implicit semantic error.
While the performance of CFG-Chain in generating the CFG of complete code is not as good as the traditional program analysis based methods, it still remains competitive.
However, CFG-Chain shows much stronger robustness in face of syntax errors and semantic deviations, compared with program analysis based methods.}
 \end{minipage}}

\begin{table}[h]
\centering
\caption{The Results of Baselines VS. Our Approach}
\vspace{-3mm}
\label{table: RQ2 result}
\begin{tabular}{|c|c|c|c|}
\hline
Dataset              & Method         & \begin{tabular}[c]{@{}c@{}}Node\\ Coverage\end{tabular} & \begin{tabular}[c]{@{}c@{}}Edge\\ Coverage\end{tabular} \\ \hline
\multirow{3}{*}{NC}  & AST-based      & 1.00                                                    & 1.00                                                    \\ \cline{2-4} 
                     & Bytecode-based & 1.00                                                    & 1.00                                                    \\ \cline{2-4} 
                     & CFG-Chain          & 0.93                                                    & 0.82                                                    \\ \hline
\multirow{3}{*}{ESE} & AST-based      & 0.64                                                    & 0.41                                                    \\ \cline{2-4} 
                     & Bytecode-based & 0                                                       & 0                                                       \\ \cline{2-4} 
                     & CFG-Chain          & 0.87                                                    & 0.80                                                    \\ \hline
\multirow{3}{*}{ISE} & AST-based      & 1.00                                                    & 0.73                                                    \\ \cline{2-4} 
                     & Bytecode-based & 1.00                                                    & 0.70                                                    \\ \cline{2-4} 
                     & CFG-Chain          & 0.93                                                    & 0.80                                                    \\ \hline
\end{tabular}
\vspace{-3mm}
\end{table}

\subsection{RQ3:The ablation study of CFG-Chain}
\vspace{2mm}
\subsubsection{Motivation}
CoT can alleviate the illusion of directly consulting LLMS, but its "epic" cues with too much accountability would make Cot-based approaches difficult to control and optimize. 
To solve this problem, we designed an AI chain. 
Step by step, the chain interacts with the LLMs to generate the CFG. 
Moreover, to improve the effectiveness of the AI chain, we also design an atomic example retrieval strategy that generates more instructive prompts.
In this RQ, we aim to investigate two aspects of our approach. Firstly, we would like to explore whether our AI chain design can effectively interact with large language models (LLMs), thus enhancing the robustness of our approach. Secondly, we would like to investigate whether the atomic example retrieval strategy could enhance the effectiveness of our AI chain.

\subsubsection{Methodology}
We set up three approach variants (CFG-D, CFG-CoT, and CFG-$Chain_{w/oAPR}$). 
Two scenarios (CFG-D vs. CFG-CoT, CFG-CoT vs. CFG-Chain) are used to evaluate the effectiveness of the AI chain.
The last one scenario (CFG-$Chain_{w/oAPR}$ vs. CFG-Chain) is used to evaluate the effectiveness of atomic examples retrieval strategy.
We use the same method as RQ2 to test the three approach variants and calculate the metric values.

\subsubsection{Result Analysis}
The experimental results are presented in Table~\ref{table: RQ3 Results}.
For CFG-D, both the node coverage and edge coverage in the three datasets are lower than that of CFG-CoT. This is because generating CFG nodes and edges directly from Java code using LLMs is challenging due to LLMs' uncertainty. For example, in Fig.~\ref{fig:Hierarchical Task Breakdown Example}(a), the two statements ``\textit{if(i+j)$<$7 System.out.println(i++)}'' may be treated as one node, instead of being treated as two separate nodes, which results in lower node coverage. In contrast, the CoT design-based prompt is more informative than that of CFG-D.

For CFG-CoT, both the node and edge coverage in the three datasets are lower than CFG-Chain, but higher than that of CFG-D. This shows that our AI chain design is superior to CoT's single-prompting approach, which completes all generative steps in a single pass using an ``epic'' prompt with hard-to-control behavior and error accumulation. In contrast, CFG-Chain breaks down the CoT into an AI chain, with each step corresponding to a separate AI unit that performs separate LLM calls. This allows CFG-Chain to interact with LLMs step by step to generate CFGs for source code.

The effect of the prompt retrieval strategy on CFG-Chain's robustness can be seen in the last three rows of Table~\ref{table: RQ3 Results}. Although CFG-$Chain_{w/oAPR}$ (without atomic prompt retrieval) produces higher code coverage and edge coverage than CFG-D and CFG-CoT, it is still inferior to CFG-Chain. The results show that the prompt retrieval strategy can increase the robustness of CFG generation.

Furthermore, we also observe that the node coverage and edge coverage of each variant are higher for the NC dataset than for the ISE dataset, and higher for the ISE dataset than for the ESE dataset. This indicates that explicit syntax errors have a greater impact on the LLM's ability to generate the nodes and edges of a CFG than implicit semantic errors.

\vspace{1mm}
\noindent\fbox{\begin{minipage}{8.2cm} \textit{Compared with directly consulting the LLm for the nodes and edges of a CFG, our AI chain design for interacting with the LLM can effectively improve the LLM's response reliability.
Our prompt retrieval strategy can further increase the robustness of CFG generation.}\end{minipage}}

\begin{table}[]
\centering
\caption{Ablation Results of CFG-Chain Variants}
\vspace{-3mm}
\label{table: RQ3 Results}
\begin{tabular}{|c|c|c|c|}
\hline
Strategies                     & Dataset & \begin{tabular}[c]{@{}c@{}}Node\\ Coverage\end{tabular} & \begin{tabular}[c]{@{}c@{}}Edge\\ Coverage\end{tabular} \\ \hline
\multirow{3}{*}{CFG-Chain}         & NC      & 0.93                                                    & 0.82                                                    \\ \cline{2-4} 
                               & ESE     & 0.87                                                    & 0.80                                                    \\ \cline{2-4} 
                               & ISE     & 0.93                                                    & 0.80                                                    \\ \hline
\multirow{3}{*}{CFG-D}   & NC      & 0.75                                                    & 0.65                                                    \\ \cline{2-4} 
                               & ESE     & 0.69                                                    & 0.51                                                    \\ \cline{2-4} 
                               & ISE     & 0.72                                                    & 0.62                                                    \\ \hline
\multirow{3}{*}{CFG-CoT} & NC      & 0.76                                                    & 0.63                                                    \\ \cline{2-4} 
                               & ESE     & 0.73                                                    & 0.61                                                    \\ \cline{2-4} 
                               & ISE     & 0.75                                                    & 0.63                                                    \\ \hline
\multirow{3}{*}{CFG-$Chain_{w/oAPR}$}   & NC      & 0.82                                                    & 0.71                                                    \\ \cline{2-4} 
                               & ESE     & 0.81                                                    & 0.64                                                    \\ \cline{2-4} 
                               & ISE     & 0.85                                                    & 0.71                                                    \\ \hline
\end{tabular}
\vspace{-5mm}
\end{table}

\section{DISCUSSION}
In this section, we summarize the principles of AI chain and prompt design patterns, and also discuss potential threats to validity. 

\subsection{Prompt Engineering Principles}
Our experiments demonstrate the need to improve the response reliability of LLMs by designing an informative CoT and breaking it down into an effective AI chain with multiple single-responsibility, composable steps. 
We summarize three AI chain principles: 1) Hierarchical Task Breakdown, which involves dividing a problem into multiple modules and submodules, and further breaking them down into functional units.
2) Unit Composition, which entails connecting functional units in a specific structure.
3) Mixing of AI Units and non-AI Units, which involves programming clear logic functional units as non-AI units, and using the LLM for functional units with fuzzy logic by designing prompts.

Prompt engineering will play an important role in problem-solving in the future.
The principles that we have outlined above can aid in designing AI chains and maximizing the potential of the LLM-based paradigm for problem-solving.

\subsection{Threats to Validity}
There are some internal and external threats to the validity of our approach. 
Firstly, manual labeling of the CFG results is a potential internal threat. To address it, we invited two annotators to label the CFG results simultaneously and measured the consistency of the results using the Kappa coefficient. The high Kappa coefficients (all higher than 80\%) indicate the reliability of the labeling results. 
Secondly, while code fixing can address some explicit syntax errors, implicit semantic errors may still exist and lead to behavioral deviation. However, our AI chain supports modular assembly, allowing us to add code fixing units to the existing AI chain and prevent the loss of behavior. 
Another potential internal threat is that we did not consider sensitive factors of the prompt, such as the number and order of examples, which may affect the results.
We plan to explore the impact of these factors in future research.

In terms of external threats, our current study only explores CFG in one statically-typed language, namely Java. To further evaluate the generalizability of our approach, we plan to investigate other statically-typed languages like C, C++, and C\#, as well as dynamic languages such as Python. 
Unlike building traditional CFG tools which demand program analysis expertise and require significant engineering and maintenance effort for different languages and their versions, extending our approach to more languages require mostly to change the prompt examples from Java to other languages.
Additionally, the emergence of new large LLMs like GPT-4~\cite{nori2023capabilities, lyu2023translating} may have an impact on our approach. 
While we are still on the GPT-4 waitlist, once we have access to it, we will utilize it to verify the effectiveness and generality of our approach.
\section{RELATED WORK}
CFG represents the order of statements and conditions for their execution, supportiing various software tasks, such as code search~\cite{guo2020graphcodebert, chen2019capturing}, code clones detection~\cite{Wang2020DetectingCC, hu2018deep, wei2017supervised}, and code classification~\cite{wang2020modular, zhang2019novel}.
The bytecode- and AST-based methods are two traditional approaches to generate CFG for java code. The former (e.g., WALA~\cite{WALA} and Soot~\cite{Soot}) converts the code into bytecode files to analyze the structure and behavior of the program at bytecode level, while the latter (e.g., Spoon~\cite{pawlak2016spoon}) generates CFG for the uncompilable code by parsing it into an AST. 
% These two approaches, however, have limitations, such as the inability to handle explicit syntax errors or implicit semantic errors caused by bad coding practices, which result in behavioral loss or deviation of CFG.
However, these traditional approaches have limitations. They may not be able to handle explicit syntax errors or implicit semantic errors that result from bad coding practices, leading to behavioral loss or deviation of CFG.
To overcome these limitations, we propose a Language Model-based (LLM) approach, which utilizes LLMs pre-trained on large amounts of code and natural language data.

LLMs (e.g., GPT-3~\cite{Brown2020LanguageMA}, CodeX~\cite{chen2021evaluating}, ChatGPT~\cite{openai_ChatGPT}), have shown significant improvements in software engineering tasks, such as requirements classification~\cite{Dong2019UnifiedLM, Luo2022PRCBERTPL, Hey2020NoRBERTTL}, FQN inference~\cite{Huang2022PrompttunedCL, Huang2022SEFK}, and code summarization~\cite{Sun2022AnEF, Gong2022SourceCS, Zhou2021AdversarialRO}. They can capture code's structural knowledge (e.g., AST~\cite{Zhang2019ANN, Diel1976LanguageRB}) and semantic knowledge (e.g., code weakness~\cite{9833571, li2018your} and API relation~\cite{Huang2023APIEA}). They can comprehend code in the same way as natural language text, preventing behavioral loss caused by explicit syntax errors, and detect implicit semantic errors based on context, avoiding behavior deviation.

Two approaches transfer LLMs to downstream tasks: supervised fine-tuning~\cite{liu2023pre, Huang2023APIEA, schick2020automatically} and in-context learning~\cite{bommasani2021opportunities, Brown2020LanguageMA, raffel2020exploring}. Supervised fine-tuning has strong few-shot learning capability by aligning downstream tasks with pre-training via prompts. However, existing supervised prompt-tuning methods cannot handle complex tasks such as CFG generation, which requires substantial data labeling. Thus, we use unsupervised in-context learning on LLMs.

In-context learning is a novel paradigm that conditions the model on task descriptions and demonstrations to generate answers for the same tasks~\cite{bommasani2021opportunities}. It has been applied to various domains, including testing~\cite{chen2022codet}, code generation~\cite{mastropaolo2023robustness}, and GUI automation~\cite{liu2022fill}.
These works use coarse-grained, direct-inquiry style prompt design.
To address complex reasoning tasks, researchers have proposed the chain of thoughts (CoT)~\cite{wang2022self, wei2022chain}.
However, existing CoT works provide only a simple instruction like ``let's do something step by step'' and cannot handle intricate tasks.
In contrast, our method is AI chain-based~\cite{wu2022ai, wu2022promptchainer,dang2022prompt}, which interacts with the model in explicit steps to generate CFGs. While the idea of AI chain has been explored for writing assistants~\cite{wu2022ai}, our AI chain involves much more complex task analysis and data flow for a domain-specific CFG generation task.

\section{CONCLUSION AND FUTURE WORK}
In this paper, we propose a novel approach for generating a behaviorally-correct CFG of statically-typed partial code by utilizing LLMs' error-tolerant and understanding ability. Our approach involves a CoT with four steps, namely \textit{structure hierarchy extraction}, \textit{nested code block extraction}, \textit{CFG generation of nested code blocks}, and \textit{fusion of all nested code blocks' CFGs}. We break down the CoT into an AI chain according to the single responsibility principle, along with effective prompt instructions, resulting in superior node and edge coverage compared to traditional program analysis based methods and the original CoT method.
Considering this performance superiority and the much lower cost to building a LLM-based CFG generation tool compared with the traditional program analysis based method, our approach provides a new LLM-based alternative solution for the development of software engineering tools that require generally significant engineering and maintenance effort.
We also provide practical principles for employing prompt engineering and AI chain in SE tasks, showcasing the potential of LLM4SE. 
By leveraging foundation models, we can focus on identifying problems for AI to solve instead of dedicating effort to data collection, labeling, model training, or program analysis.

\normalem

\bibliographystyle{unsrt}
\bibliography{sample-base}
\end{document}